\documentclass[preprintnumbers,twocolumn,prl,aps,superscriptaddress,footinbib,amsfonts,amsmath,amssymb,showpacs]{revtex4-1}

\usepackage[T1]{fontenc}
\usepackage{blindtext}

\usepackage{graphics,graphicx}
\usepackage{dcolumn}
\usepackage{bm}
\usepackage{epsfig}
\usepackage[usenames]{color}
\usepackage[hidelinks]{hyperref} 
\usepackage{mathbbol}
\usepackage{epstopdf}
\usepackage{simplewick}
\usepackage[utf8]{inputenc} 
\usepackage{slashed}
\usepackage{soul}
\usepackage[dvipsnames]{xcolor}
\usepackage[export]{adjustbox}

\usepackage{multirow}

\usepackage{lineno} 

\usepackage{hyperref} 
\hypersetup{
    colorlinks,
    citecolor=black,
    filecolor=black,
    linkcolor=black,
    urlcolor=black
}

\newcommand{\nn}{\nonumber}

\newcommand{\ot}{\leftarrow}

\newcommand{\be}{\begin{equation}}
\newcommand{\ee}{\end{equation}}

\newcommand{\ba}{\begin{eqnarray}}
\newcommand{\ea}{\end{eqnarray}}

\def\bea#1\eea{\begin{align}#1\end{align}}

\usepackage[normalem]{ulem}

\newcommand{\new}[1]{{\color{blue}#1}}

\begin{document}

\preprint{JLAB-THY-20-3296}

\title{N$^3$LO extraction of the Sivers function from SIDIS, Drell-Yan, and $W^\pm/Z$ data}

\newcommand*{\REG}{Institut f\"ur Theoretische Physik, Universit\"at Regensburg, D-93040 Regensburg, Germany}\affiliation{\REG}
\newcommand*{\PSU}{Division of Science, Penn State University Berks, Reading, Pennsylvania 19610, USA}\affiliation{\PSU}
\newcommand*{\JLAB}{Thomas Jefferson National Accelerator Facility, Newport News, VA 23606, USA}\affiliation{\JLAB}

\author{Marcin~Bury}\email{marcin.bury@ur.de}\affiliation{\REG}
\author{Alexei Prokudin}\email{prokudin@jlab.org}\affiliation{\PSU}\affiliation{\JLAB}
\author{Alexey~Vladimirov}\email{alexey.vladimirov@ur.de}\affiliation{\REG}

\begin{abstract}
\noindent
We perform the global analysis of polarized Semi-Inclusive Deep Inelastic Scattering (SIDIS),  pion-induced polarized Drell-Yan (DY), and  $W^\pm/Z$ boson production data and extract the Sivers function for $u$, $d$, $s$ and for sea-quarks. We use the framework of transverse momentum dependent factorization at N$^3$LO accuracy. The Qiu-Sterman function is determined in a model-independent way from the extracted Sivers function. We also evaluate the significance of the predicted sign change of Sivers function in DY with respect to SIDIS. 
\end{abstract}

\pacs{}
\maketitle

{\bf Introduction.}
The three-dimensional (3D) hadron structure is an important topic of theoretical, phenomenological, and experimental studies in nuclear physics.  In the momentum space, the 3D nucleon structure is described in terms of  Transverse Momentum Dependent distributions and fragmentation functions, collectively called TMDs, which depend both on the collinear momentum fraction and the transverse momentum of parton. TMD factorization theorem \cite{Collins:2011zzd,GarciaEchevarria:2011rb} provides consistent operator definition and  evolution of TMDs. Among TMDs, the Sivers function $f_{1T}^\perp(x, k_T)$~\cite{Sivers:1989cc,Sivers:1990fh} is the most intriguing since it describes distribution of unpolarized quarks inside a transversely polarized  nucleon and generates  single-spin asymmetries (SSAs). 

The Sivers function arises from interaction of the initial or final state quark with the remnant of the nucleon and thus, many of its features reveal the gauge link structure that reflects the kinematics of the underlining process \cite{Belitsky:2002sm}. Above all, the difference between initial an final state gauge contours leads to  the opposite signs for Sivers functions in SIDIS and DY kinematics \cite{Brodsky:2002rv,Brodsky:2002cx,Collins:2002kn}
\begin{eqnarray}\label{eq:sign}
f_{1T}^\perp(x,k_T)_{\text{[SIDIS]}}=-f_{1T}^\perp(x,k_T)_{\text{[DY]}}.
\end{eqnarray}
In the limit of the large transverse momentum the Sivers function is related~\cite{Ji:2006ub} to the key ingredient of collinear factorization of SSAs, the  Qiu-Sterman (QS) function~\cite{Efremov:1981sh,Efremov:1983eb,Qiu:1991pp,Qiu:1998ia}, which describes the correlation of quarks with the null-momentum gluon field. Therefore, the measurement of Sivers function and the exploration of its properties is a crucial test of our understanding of the strong force, and one of the goals of polarized SIDIS and DY experimental programs of  future and existing experimental facilities such as the Electron Ion Collider~\cite{Boer:2011fh,Accardi:2012qut}, Jefferson Lab 12~GeV Upgrade~\cite{Dudek:2012vr}, RHIC~\cite{Aschenauer:2015eha} at BNL, COMPASS~\cite{Gautheron:2010wva,Bradamante:2018ick} at CERN.

In this work, we perform the global analysis of transverse spin asymmetries at next-to-next-to-next-to-leading order (N$^3$LO) perturbative precision within TMD factorization approach and extract Sivers function. Several important features make our results stand out from the previous results~\cite{Efremov:2004tp,Vogelsang:2005cs,Anselmino:2005ea,Anselmino:2005an,Collins:2005ie,Anselmino:2008sga,Anselmino:2010bs,Bacchetta:2011gx,Gamberg:2013kla,Sun:2013dya,Echevarria:2014xaa,Anselmino:2016uie,Boglione:2018dqd,Bacchetta:2020gko,Echevarria:2020hpy,Cammarota:2020qcw}. First of all, we use unprecedented N$^3$LO perturbative precision, together with the $\zeta$-prescription \cite{Scimemi:2018xaf}. Secondly, we use unpolarized proton and pion TMDs extracted from the global fit of SIDIS and DY data \cite{Scimemi:2019cmh,Vladimirov:2019bfa} at the same perturbative order and scheme, which allows us for the first time to consistently describe SIDIS and DY, $W^\pm/Z$ experimental data. Lastly, we use the novel model-independent approach to obtain QS function from the Sivers function. Also, we estimate the significance of sign flip relation (\ref{eq:sign}). 

{\bf SIDIS process.} The most precise experimental measurement related to the Sivers function comes from SIDIS on a transversely polarized target ($e(l)+h_1(P,S)\to e(l')+h_2(P_h)+X$). The relevant part of the cross section has the following structure \cite{Gourdin:1973qx,Kotzinian:1994dv,Diehl:2005pc,Bacchetta:2006tn}
\begin{eqnarray}
\label{def:SIDIS}
\frac{d\sigma}{d \mathcal{PS}}
=
\sigma_0
\left\{
F_{UU ,T} 
+|S_\perp| \sin(\phi_h-\phi_S) F_{UT ,T}^{\sin(\phi_h -\phi_S)}  
\right\}
\end{eqnarray}
where $d \mathcal{PS}$ $=$ $dx \, dy\, d\psi \,dz\, d\phi_h\, d P_{hT}^2$
and $\sigma_0$ $\equiv$ ${\alpha_{\text{em}}^2(Q) y}/{(2(1-\varepsilon)Q^2)}$ and the usual DIS variables are used \cite{Bacchetta:2006tn}.  The variable $P_{hT}$ is the  transverse  momentum of the produced hadron $h_2$ in the laboratory frame. The azimuthal angle $\phi_h$ and $\phi_S$ are measured relative to the lepton plane~\cite{Bacchetta:2004jz}. The single-spin  Sivers asymmetry in SIDIS is defined as the ratio of structure functions and can be written in the TMD factorization as
\begin{eqnarray}\label{eq:sidisaut}
A_{UT}^{\sin(\phi_h-\phi_S)}&\equiv&\frac{F_{UT ,T}^{\sin(\phi_h -\phi_S)}}{F_{UU ,T}} = -M\frac{\mathcal{B}^{\text{SIDIS}}_1\left[f_{1T}^\perp D_{1}\right]}{\mathcal{B}^{\text{SIDIS}}_0\left[f_{1} D_{1}\right]}\; .
\end{eqnarray}
where $M$ is the mass of the nucleon $h_1$, and 
\begin{align}
\label{eq:notation}
    \mathcal{B}^{\text{SIDIS}}_n[f D] &\equiv \sum_{q}e_q^2 \int_0^\infty \frac{b db}{2\pi} b^n J_n\left(\frac{b |P_{hT}|}{z}\right) \nonumber \\ & \times f_{q\ot h_1}(x,b;\mu,\zeta_1)D_{q\to h_2}(z,b;\mu,\zeta_2)
\end{align}
where $f$ and $D$ are TMD parton distribution function (PDF) and fragmentation function (FF), $J_n$ is the Bessel function of the first kind and the summation runs over all active quarks and anti-quarks $q$ with electric charge $e_q$. The arguments $\mu$ and $\zeta$ are the ultra-violet and the rapidity renormalization scale, correspondingly. The $Q$-dependence of the ratio in Eq.~(\ref{eq:sidisaut}) is due to the scales $\zeta_{1,2}$, which obey $\zeta_1\zeta_2=Q^4$  \cite{Collins:2011zzd,Vladimirov:2017ksc}.  
To respect it, we fix $\zeta_1=\zeta_2=Q^2$, and also $\mu^2=Q^2$. The dependence on $(\mu,\zeta)$ of a TMD distribution is dictated by the pair of TMD evolution equations \cite{Collins:2011zzd,Scimemi:2018xaf}, which, in turn, relate measurements made at different energies. In this work we use the $\zeta$-prescription \cite{Scimemi:2018xaf} which consists in selecting the reference scale $(\mu,\zeta)=(\mu,\zeta_\mu(b))$ on the equipotential line of the field anomalous dimension that passes through the saddle point. In this case, the reference TMD distribution is independent on $\mu$ (by definition) and perturbatively finite in the whole range of $\mu$ and $b$.  The solution of the evolution equations can be written \cite{Scimemi:2017etj,Scimemi:2018xaf} in the following simple form
\begin{align}
f_{1T,q\ot h}^\perp(x,b;\mu,\zeta)=\left(\frac{\zeta}{\zeta_\mu(b)}\right)^{-\mathcal{D}(b,\mu)}f_{1T,q\ot h}^\perp(x,b),
\label{eq:optimal}
\end{align}
and similar for other TMDs. The function $f_{1T,q\ot h}^\perp(x,b)=f_{1T,q\ot h}^\perp(x,b;\mu,\zeta_\mu(b))$ on the right-hand side of Eq.~\eqref{eq:optimal} is the optimal Sivers function~\cite{Scimemi:2017etj}. The function $\zeta_\mu(b)$ is a calculable function of the universal non-perturbative Collins-Soper kernel $\mathcal{D}(b,\mu)$~\footnote{Our definition of the rapidity anomalous dimension corresponds to $\tilde K$ and $\gamma_\nu$ used in Refs.\cite{Collins:2011zzd} and \cite{Chiu:2011qc} as $\mathcal{D} = -\tilde K/2=-\gamma_\nu/2$.}. The N$^3$LO expression used in this work is given in Ref.~\cite{Scimemi:2019cmh}.

{\bf Drell-Yan process.} The relevant part of the differential cross-section for DY reaction $(h_1(P_1,S)+h_2(P_2)\to l^+(l)+l^-(l')+X$) is \cite{Arnold:2008kf}
\begin{eqnarray}\label{def:DY}
&& \frac{d\sigma}{d \mathcal{PS}}=\sigma_0^{[DY]}\left\{
F_{UU}^1+|S_T|\sin(\varphi-\phi_S) F_{TU}^1
 \right\},
\end{eqnarray}
where $d \mathcal{PS} = dQ^2 \, dy \, d\varphi \, dq_T^2$, $\sigma_0^{[DY]} = {\alpha_{\text{em}}^2(Q)}/{(9 s Q^2)}$. The variables $\varphi$ and $q_T$ are the angle and the transverse  momentum of  the electro-weak boson  measured in the center-of-mass frame and $y$ is its rapidity. The experimentally measured transverse spin asymmetry is
\begin{eqnarray}\label{th:AUT}
A_{TU}\equiv\frac{F_{TU}^1}{F_{UU}^1}=
-M\frac{\mathcal{B}^{\text{DY}}_1[f_{1T}^\perp \, f_1]}{\mathcal{B}^{\text{DY}}_0[f_1\, f_1]}\, ,
\end{eqnarray}
where $M$ is the mass of the polarised hadron $h_1$, and
\begin{align}
\label{eq:notationdy}
\mathcal{B}^{\text{DY}}_n[f_1 \, f_2] &\equiv \sum_{q}e_q^2 \int_0^\infty \frac{b db}{2\pi} b^n J_n\left(b |q_T|\right) \nonumber \\ &\times f_{1; q\ot h_1}(x_1,b;\mu,\zeta_1)f_{2; \bar q\ot h_2}(x_2,b;\mu,\zeta_2)
\end{align}
where $f_1$ and $f_2$ are TMD PDFs for hadrons $h_1$ and $h_2$. 

Often, the experiment provides measurements related to $A_{TU}$ (\ref{th:AUT}). In particular, the process $h_1(P_1)+h_2(P_2,S)\to l^+l^-+X$ (i.e. with the polarized hadron $h_2$) measured by COMPASS~\cite{Aghasyan:2017jop} is described by $A_{UT}=-A_{TU}(f_{1T}^\perp\leftrightarrow f_1)$, where the exchange of Sivers and unpolarized TMD PDFs takes place in the numerator of (\ref{th:AUT}) and $M$ refers to $h_2$.  Another important case is the asymmetry $A_N$~\cite{Adamczyk:2015gyk} measured by STAR Collaboration and defined such that $A_N=-A_{TU}$ \cite{Kang:2009sm}. The STAR measurements are made for $W^\pm/Z$-boson production, and thus $\mathcal{B}^{\text{DY}}_n$ (\ref{eq:notationdy}) should be updated replacing $\sum_q e_q^2$ by an appropriates structure, which can be found e.g. in Ref. \cite{Scimemi:2019cmh}.

{\bf Non-perturbative input.} In addition to the Sivers function, SSAs (\ref{eq:sidisaut},\ref{th:AUT}) contain non-perturbative unpolarized TMDs and the Collins-Soper kernel. We use these functions from Ref. \cite{Scimemi:2019cmh}  (\texttt{SV19}). \texttt{SV19} was made by the global analysis of a large set of DY and SIDIS data, including precise measurements made by the LHC, performed with N$^3$LO TMD evolution and NNLO matching to the collinear distributions. The unpolarized TMD PDFs for the pion were extracted in the same framework in Ref.~\cite{Vladimirov:2019bfa}. In these extractions the Collins-Soper kernel is parameterized as
\begin{eqnarray}\label{def:RAD}
\mathcal{D}(b,\mu)=\mathcal{D}_{\text{resum}}(b^*,\mu)+c_0 bb^*,
\end{eqnarray}
where $b^*=b/\sqrt{1+\left(b/(2\; {\rm {GeV}^{-1}})\right)^{2}}$, $\mathcal{D}_{\text{resum}}$ is the resummed N$^3$LO expression for the perturbative part \cite{Vladimirov:2016dll}, and $c_0$ is a free parameter. The linear behavior at large-$b$ of Eq.~(\ref{def:RAD}) is in agreement with the predicted non perturbative behavior  ~\cite{Collins:2014jpa,Vladimirov:2020umg} and coefficient $c_0$ can be related to the gluon-condensate~\cite{Vladimirov:2020umg}.

It is customary in the TMD phenomenology to match TMDs to collinear distributions at small-$b$ \cite{Collins:2011zzd,Aybat:2011zv,Echevarria:2016scs,Scimemi:2019gge}. In the present work, we do not use the matching of the Sivers to QS function \cite{Scimemi:2019gge,Sun:2013dya,Dai:2014ala}, since it is not  beneficial in the Sivers case. The reason is that QS function is not an autonomous function, but mixes with other twist-3 distributions \cite{Braun:2009mi}. Therefore, a consistent implementation of the matching requires introduction of several unknown functions -- subjects of fitting. Instead, we use the reversed procedure. We consider the optimal Sivers function as a generic non-perturbative function that is extracted directly from the data. QS function is then obtained from the small-$b$ limit of the extracted Sivers function. For the Sivers function, we use the following ansatz
\begin{eqnarray}\label{def:model}
f_{1T;q\ot h}^\perp(x,b)&=N_q \frac{(1-x) x^{\beta_q} (1+\epsilon_q x)}{n(\beta_q,\epsilon_q)} \nn \\
&\times \exp\left(-\frac{r_0+x r_1}{\sqrt{1+r_2 x^2 b^2}}b^2\right),
\end{eqnarray}
where $n(\beta,\epsilon)= (3+\beta+\epsilon+\epsilon \beta)\Gamma(\beta+1)/\Gamma(\beta+4)$, such that
\begin{eqnarray}
\int_0^1 dx f_{1T;q\ot h}^\perp(x,0)=N_q.
\end{eqnarray}
We will distinguish separate functions for $u$, $d$, $s$ quarks, and a single \textit{sea} Sivers function for $\bar u$, $\bar d$ and $\bar s$ quarks. The Sivers function does not have the probabilistic interpretation and can have nodes~\cite{Bacchetta:2003rz}, which is realized by the parameter $\epsilon$. We set $\beta_s=\beta_{sea}$ and $\epsilon_s=\epsilon_{sea}=0$, since they are not restricted by the current experimental data. In total, we have 12 free parameters in our fit.

Notice that the absence of the small-$b$ matching is advantageous for our analysis as it allows both to circumvent the difficulties of evolution of QS functions and to reach N$^3$LO precision. Such a strategy is allowed in the $\zeta$-prescription, and would also work in other fixed scale prescriptions~\cite{Collins:2014jpa}, but is not consistent in the resummation-like schemes e.g. used in Refs~ \cite{Echevarria:2014xaa,Bacchetta:2020gko,Echevarria:2020hpy}.

{\bf Fit of the data.}
The TMD factorization theorems are derived in the limit of large-$Q$ and a small relative transverse momentum $\delta$, defined as
$\delta={|P_{hT}|}/{(zQ)}$ in SIDIS, $\delta={|q_T|}/{Q}$ in DY.
We apply the following selection criteria~\cite{Scimemi:2019cmh,Vladimirov:2019bfa} onto the experimental data
\begin{eqnarray}\label{def:cuts}
\langle Q\rangle >2\; \text{GeV}\qquad \text{and}\qquad \delta<0.3.
\end{eqnarray}

\begin{figure}[t]
\begin{center}
\includegraphics[width=0.49\textwidth]{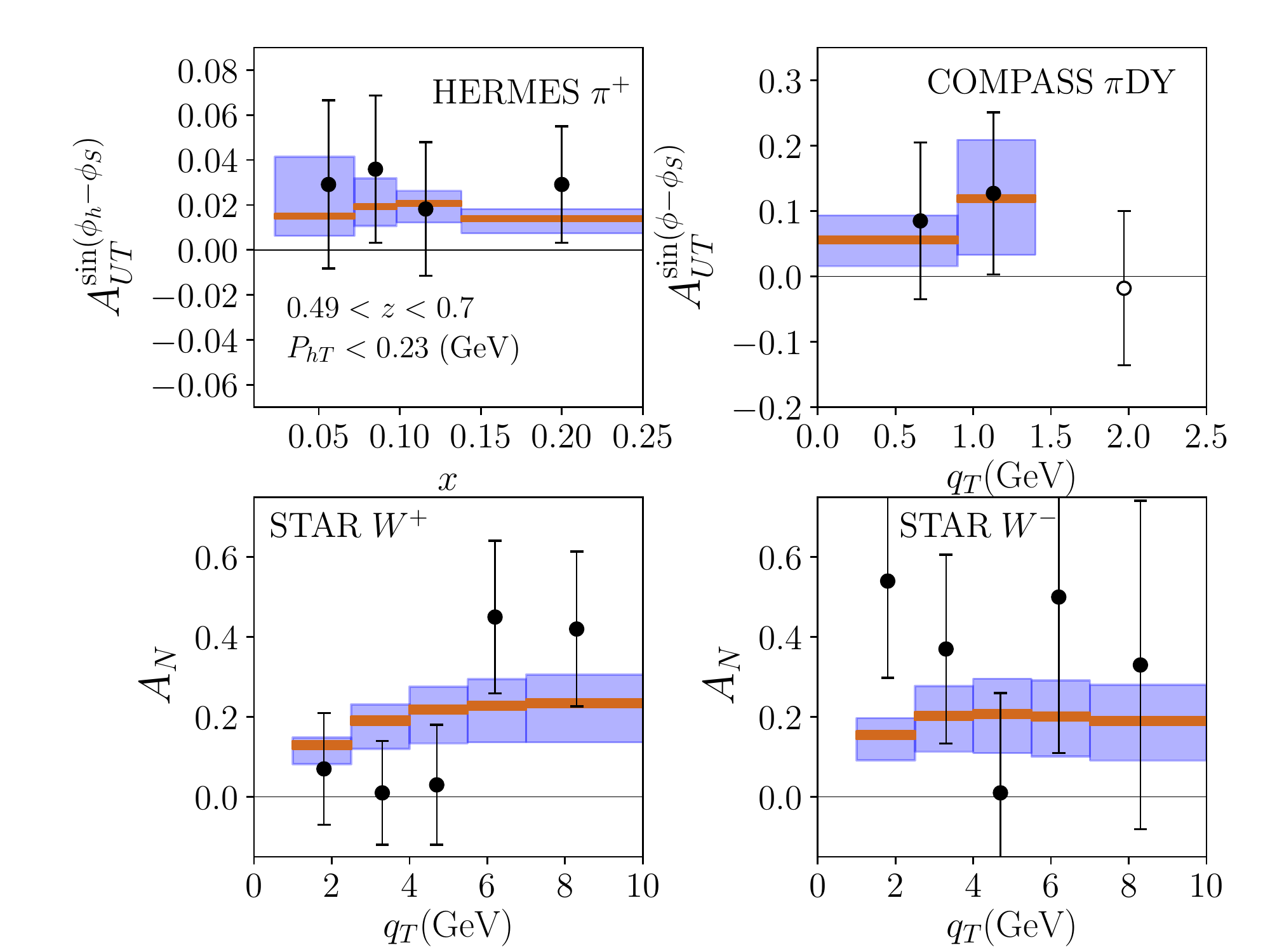}
\end{center}
\vskip -0.5cm
\caption{\label{fig:ANqt} Examples of data description of SIDIS+DY N$^3$LO fit for HERMES SIDIS~\cite{Airapetian:2020zzo}, COMPASS pion-induced DY~\cite{Aghasyan:2017jop} and STAR $W^\pm/Z$ data~\cite{Adamczyk:2015gyk}. Open symbols: data not used in the fit. Orange line is the CF and the blue box is 68\%CI. }
\end{figure}

 The Sivers asymmetry has been measured in SIDIS and DY \cite{Airapetian:2009ae,Airapetian:2020zzo,Alekseev:2008aa,Adolph:2014zba,Qian:2011py,Aghasyan:2017jop,Adamczyk:2015gyk}. In total, after data selection cuts (\ref{def:cuts}), we use 76 experimental points. We have 63 points from SIDIS measurements collected in $\pi^\pm$ and $K^\pm$ production off polarized proton target at  HERMES~\cite{Airapetian:2020zzo}, off deuterium target from COMPASS~\cite{Alekseev:2008aa}, and $^3$He target from JLab~\cite{Qian:2011py,Zhao:2014qvx}, $h^\pm$ data on the proton target  from COMPASS~\cite{Adolph:2016dvl}. We use 13 points from DY measurements of $W^\pm/Z$ production from STAR~\cite{Adamczyk:2015gyk} and pion-induced DY from COMPASS~\cite{Aghasyan:2017jop}. Let us emphasize that the recent 3D binned data~\cite{Airapetian:2020zzo} from  HERMES allowed us to select sufficient number of data (46 points) from SIDIS measurements. COMPASS and JLab  measurements in SIDIS  are done by projecting the same data onto $x$, $z$, and $P_{hT}$. In order not to use the same data multiple times and for better adjustment to TMD factorization limit, we use only $P_{hT}$-projections. 

The evaluation of the theory prediction for a given set of model parameters is made by \texttt{artemide} \cite{artemide}. The estimation of uncertainties utilizes the replica method~\cite{Ball:2008by}, which consists of the fits of data replicas generated in accordance with experimental uncertainties. From the obtained distribution of 500 replicas, we determine the values and the errors on parameters and observables, including, for the first time, propagation of the errors due to the unpolarized TMDs. We use the mean value of the resulting distributions due to \texttt{SV19} uncertainty as the central fit value (CF value), which is our best estimate of the true values for the free parameters. The uncertainty is given by a 68\% confidence interval (68\%CI) is computed by the bootstrap method. The resulting  replicas are available as a part of \texttt{artemide} \cite{artemide_sivers}. 

We performed several fits with different setups. In particular, we distinguish the fits with and without the inclusion of DY data. We found that the Sivers function extracted in SIDIS-only fit nicely describes the DY data without extra tuning. Indeed, N$^3$LO SIDIS-only fit has  $\chi^2/N_{pt} = 0.87$ and without any adjustment describes also DY data with $\chi^2/N_{pt} = 1.23$. 

The combined SIDIS+DY fit reaches a very good overall $\chi^2/N_{pt} = 0.88$ for all 76 DY and SIDIS data points, with $\chi^2/N_{pt}=0.88$ for SIDIS and $\chi^2/N_{pt}=0.90$ for DY. Parameters of Sivers function resulting from  SIDIS-only and SIDIS+DY fits are compatible with each other \new{\footnote{The numerical values of parameters and other subsidiary information will be presented in the following publication.}}.  The quality of data description in SIDIS+DY N$^3$LO fit can be seen in Fig.~\ref{fig:ANqt}.
\begin{figure}[t]
\begin{center}
\includegraphics[width=0.34\textwidth]{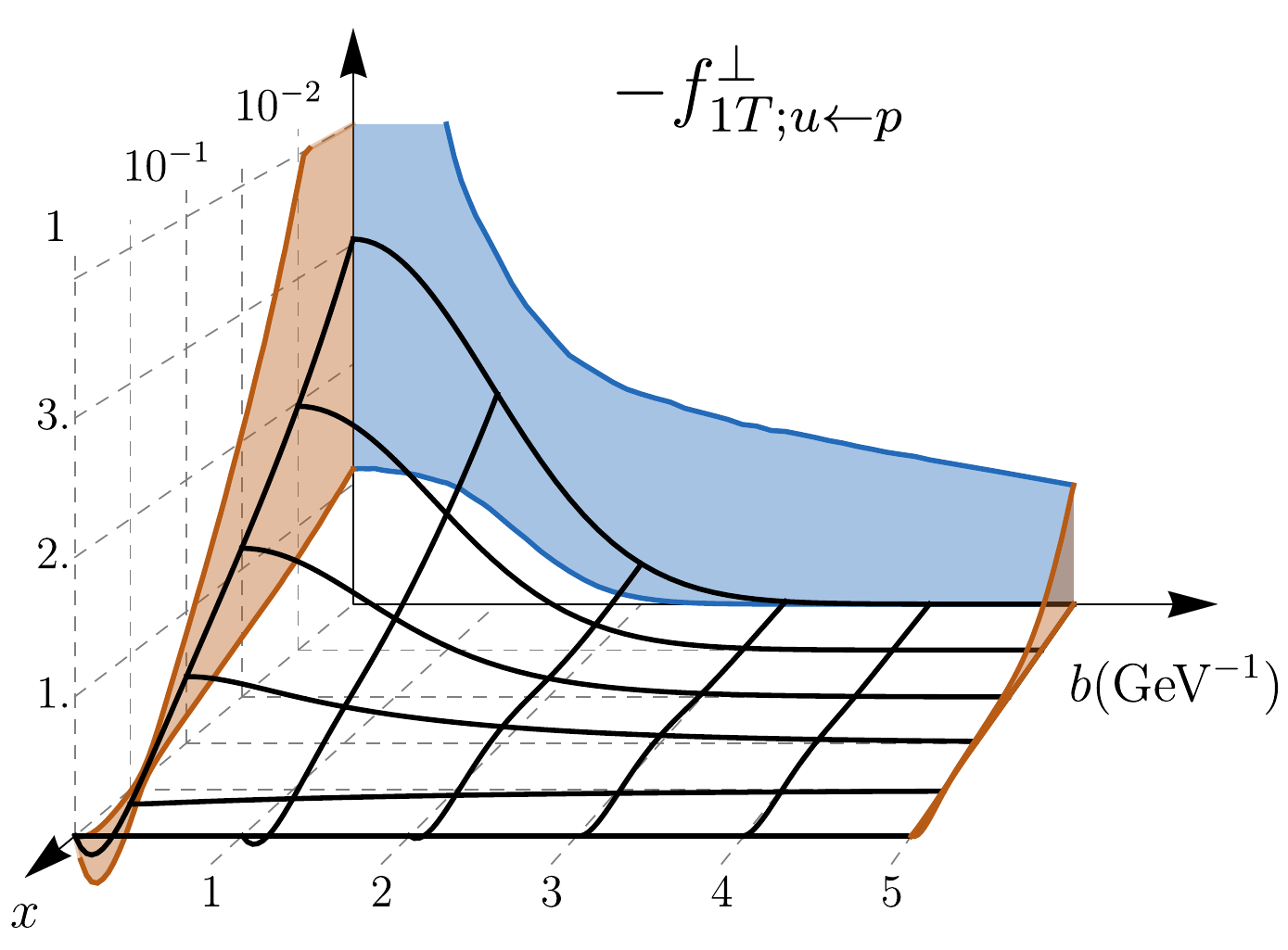}(a)
\includegraphics[width=0.34\textwidth]{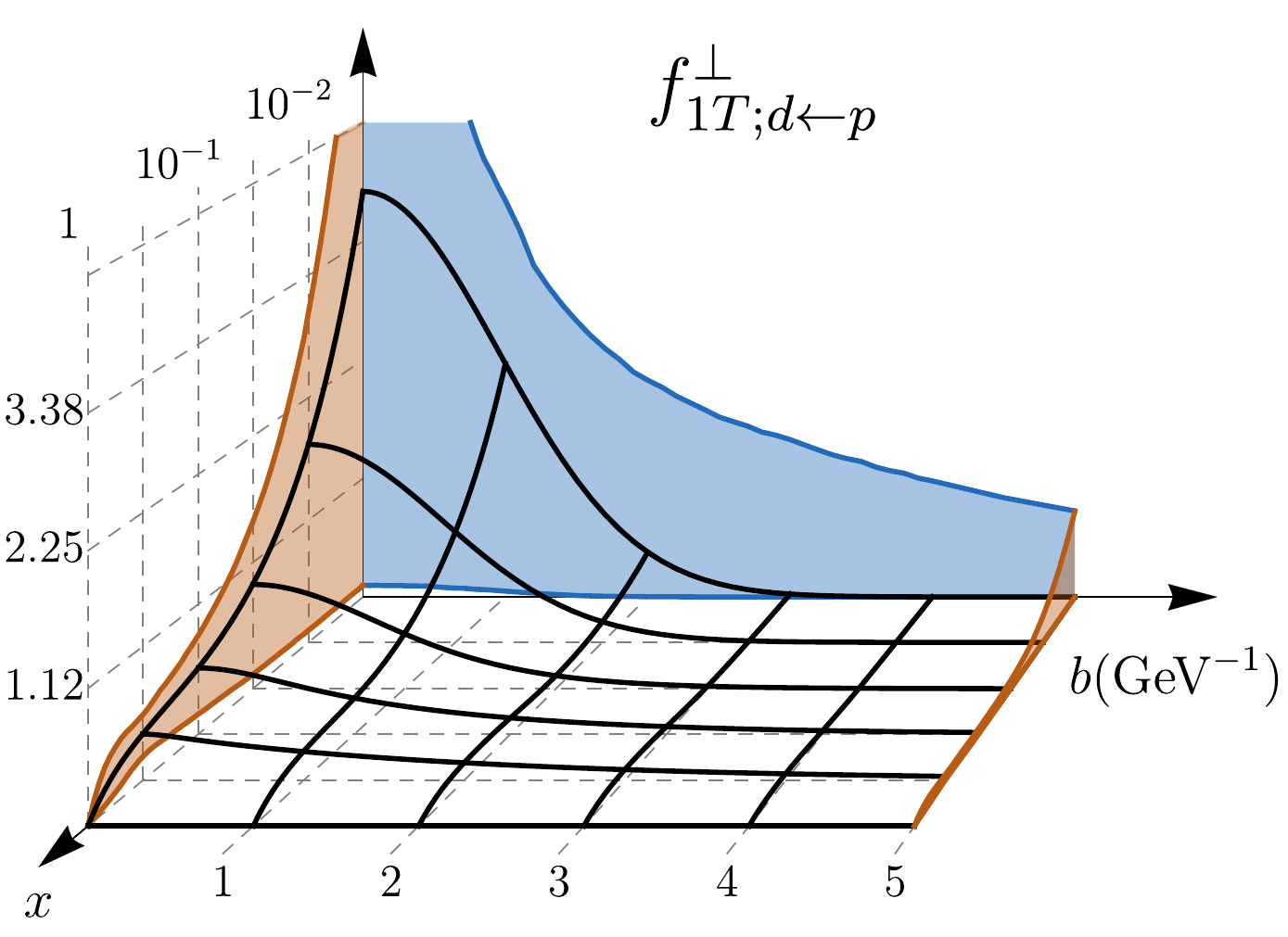}(b)
\end{center}
\caption{\label{fig:sofaPlot} The three-dimensional $(b,x)$-landscape of the optimal Sivers function $f_{1T;q\ot p}^\perp(x,b)$ for $u$-quark (a) and $d$-quark (b). The grid shows the CF value, whereas the shaded (blue and brown) regions on the boundaries demonstrate the 68\%CI.}
\end{figure}

We have performed a fit without the sign change of Sivers function from Eq.~(\ref{eq:sign}) in order to estimate the significance of the sign change from the data. The resulting fit does exhibit tensions between DY and SIDIS data sets, however, the fit has $\chi^2/N_{pt} = 1.0$ and cannot exclude the same sign of Sivers functions in DY and SIDIS. The sign of the sea-quark Sivers function plays here the central role. Indeed, the sign of DY cross-section is mostly determined by the sea-contribution due to favored $q+\bar q\to W/Z/\gamma$ sub-process, whereas the sea-contribution in SIDIS is suppressed. Therefore, with the current data precision, the flip of the sign for $N_{\text{sea}}$ parameter alone is sufficient to describe the data and almost compensates the effect of the overall sign-flip (\ref{eq:sign}) at the level of the cross-section. The future data from RHIC and COMPASS together with EIC and JLab will allow us to establish the confirmation of the sign change (\ref{eq:sign}).

{\bf Extracted Sivers functions} The extracted  Sivers functions in $b$-space for $u$ and $d$-quarks are shown in Fig.~\ref{fig:sofaPlot}. One can see that our results confirm the signs of $u$-quark (negative) and $d$-quark (positive) at middle-$x$ range known from the previous analyses~\cite{Efremov:2004tp,Vogelsang:2005cs,Anselmino:2005ea,Anselmino:2005an,Collins:2005ie,Anselmino:2008sga,Anselmino:2010bs,Bacchetta:2011gx,Gamberg:2013kla,Sun:2013dya,Echevarria:2014xaa,Anselmino:2016uie,Bacchetta:2020gko,Echevarria:2020hpy,Cammarota:2020qcw}, and also shows a node for $u$-quark at large-$x$. We have not explicitly used the positivity relation~\cite{Bacchetta:1999kz} of Sivers functions because it is only a LO statement and can be violated in higher order calculations. However, we verified  numerically that our results do not exhibit any substantial violation of positivity bounds.

\begin{figure}[t]
\begin{center}
\includegraphics[width=0.49\textwidth]{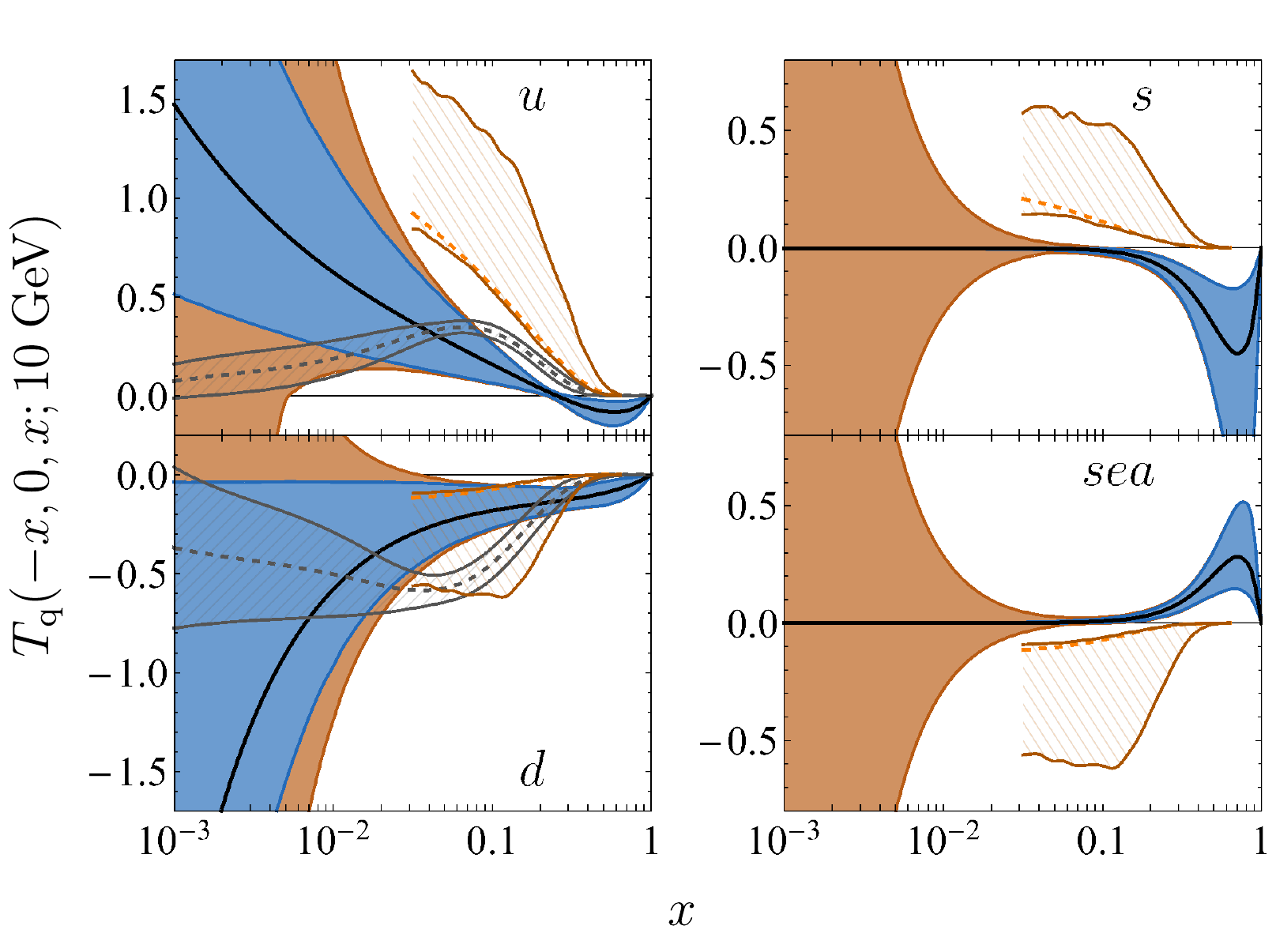}
\vskip -0.2cm
\end{center}
\caption{\label{fig:QS} Qiu-Sterman function at $\mu=10$ GeV for different quark flavors, derived from the Sivers function via Eq.~(\ref{th:QS=f}). The black line shows the CF value and blue band shows 68\%CI. The brown band shows the band obtained by adding the gluon contribution $G^{(+)}$. We compare our results to JAM20 \cite{Cammarota:2020qcw} (gray dashed lines) and 
ETK20~\cite{Echevarria:2020hpy} (orange dashed lines).} 
\end{figure}

The magnitude of $s$ and $sea$ quarks contribution in our fit is substantially different from other extractions where the the biased anzatz $f_{1T}^\perp(x)\propto f_1(x)$ is used~\cite{Anselmino:2005ea,Collins:2005ie,Anselmino:2008sga,Anselmino:2010bs,Bacchetta:2011gx,Gamberg:2013kla,Sun:2013dya,Echevarria:2014xaa,Anselmino:2016uie,Bacchetta:2020gko,Echevarria:2020hpy} and the non-valence contribution is artificially suppressed. In our case, the sea- and $s$-quark Sivers functions are comparable in size with $u$ and $d$-quarks, at $x>0.1$ (and vanish at $x<0.1$). Our unbiased extraction of the Sivers function reproduces large SSA measured in the DY $W^\pm/Z$ processes, see Fig.~\ref{fig:ANqt}.

{\bf Determination of the Qiu-Sterman function}
The  Sivers function at small-$b$ can be expressed via the operator product expansion (OPE) through the twist-3 distributions  \cite{Scimemi:2019gge,Sun:2013hua,Dai:2014ala}. At the OPE scale $\mu=\mu_b\equiv 2 \exp(-\gamma_E)/b$ the NLO matching expression \cite{Scimemi:2019gge} depends only on QS function and can be inverted. We obtain the following relation
\begin{widetext}
\begin{eqnarray}\label{th:QS=f}
&T_q(-x,0,x;\mu_b)=-\frac{1}{\pi}f_{1T;q\ot h}^\perp(x,b)-\frac{\alpha_s(\mu_b)}{4\pi^2} \int\limits_{x}^{1} \frac{dy}{y} 
\Big[
\frac{\bar y}{N_c}f_{1T;q\ot h}^\perp\left(\frac{x}{y},b\right)+
 \frac{3y^2\bar{y}}{2x}G^{(+)}\left(-\frac{x}{y},0,\frac{x}{y};\mu_b\right)\Big]
+\mathcal{O}(a_s^2,b^2)\; ,
\end{eqnarray}
\end{widetext}
where $\bar{y} = 1-y$,  $\alpha_s$ is the strong coupling constant, $T_q$ and $G^{(+)}$ are QS quark and gluon functions. 
This expression is valid only for small (non-zero) values of $b$. We use $b\simeq 0.11$ GeV$^{-1}$ such that $\mu_b = 10$ GeV. The resulting QS-functions are shown in Fig.~\ref{fig:QS}. To estimate the uncertainty due to the unknown gluon contribution we allow for $G^{(+)}=\pm (|T_u| + |T_d|)$. The resulting 68\%CI uncertainty band and comparisons to Refs.~\cite{Cammarota:2020qcw,Echevarria:2020hpy} are also shown in Fig.~\ref{fig:QS}.

{\bf Conclusions.}
In this paper, we have performed the first extraction of the Sivers function that consistently utilizes previously extracted unpolarised proton and pion TMDs, and uses SIDIS, pion-induced Drell-Yan, and $W^\pm/Z$-bozon production experimental data. The extraction is performed at unprecedented N$^3$LO perturbative precision within the $\zeta$-prescription that allows us to unambiguously relate the Sivers function and QS function. This relation has been used to obtain QS function and to evaluate the influence of the unknown gluon QS function. We also examined the significance of the predicted sign change of Sivers functions in SIDIS and DY processes. Our results compare well in  magnitude with the existing extractions~\cite{Efremov:2004tp,Vogelsang:2005cs,Anselmino:2005ea,Anselmino:2005an,Collins:2005ie,Anselmino:2008sga,Anselmino:2010bs,Bacchetta:2011gx,Gamberg:2013kla,Sun:2013dya,Echevarria:2014xaa,Anselmino:2016uie,Boglione:2018dqd,Cammarota:2020qcw,Bacchetta:2020gko,Echevarria:2020hpy} and confirm the signs of Sivers functions for $u$ and $d$ quarks while we also obtain non negligible Sivers functions for anti-quarks. 

Our results set a new benchmark and the standard of precision for studies of TMD polarized functions and are going to be important for theoretical, phenomenological, and experimental studies of the 3D nucleon structure and for the planning of experimental programs.

{\bf Acknowledgments.}
The authors are thankful to Gunar Schnell and Bakur Parsamyan for discussions of experimental data, and to John Terry for sharing results of ETK20~\cite{Echevarria:2020hpy}. 
This work was partially supported by 
DFG FOR 2926 ``Next Generation pQCD for  Hadron  Structure:  Preparing  for  the  EIC'',  project number 430824754 (M.B and A.V) and by the National Science Foundation under the Contract  No.~PHY-2012002 (A.P.), and by the US Department of Energy under contract No.~DE-AC05-06OR23177 (A.P.) under which JSA, LLC operates Jefferson Lab, and within the framework of the TMD Topical Collaboration (A.P.).

\bibliography{biblio}

\end{document}